\documentclass[a4paper,fleqn,usenatbib]{mnras}

\usepackage[T1]{fontenc}
\usepackage{ae,aecompl}
\usepackage[final]{changes}
\colorlet{Changes@Color}{purple}

\usepackage{graphicx}	
\usepackage{amsmath}	
\usepackage{amssymb}	
\usepackage{subfigure}

\title[Variability of the Lowest Mass Objects in AB Doradus]{Variability of the Lowest Mass Objects in the AB Doradus Moving Group}

\author[J. M. Vos et al.] 
{Johanna M. Vos,$^{1,2}$\thanks{E-mail: jvos@roe.ac.uk}
Katelyn N. Allers,$^{3}$ 
Beth A. Biller,$^{1,2}$ 
Michael C. Liu,$^{4}$ 
\newauthor Trent J. Dupuy,$^{5}$ 
 Jack F. Gallimore,$^{3}$ 
Dunni Adenuga$^{3}$
and William M. J. Best$^{4}$ 
\\
$^{1}$SUPA Institute for Astronomy, University of Edinburgh, Blackford Hill View, Edinburgh EH9 3HJ, UK\\
$^{2}$Centre for Exoplanet Science, University of Edinburgh, UK\\
$^{3}$Department of Physics and Astronomy, Bucknell University, Lewisburg, PA 17837, USA\\
$^{4}$Institute for Astronomy, University of Hawaii, 2680 Woodlawn Drive, Honolulu, HI 96822, USA\\
$^{5}$The University of Texas at Austin, Department of Astronomy, 2515 Speedway C1400, Austin, TX 78712, USA\\
}

\date{Accepted XXX. Received YYY; in original form ZZZ}

\pubyear{2017}

\begin{document}
\label{firstpage}
\pagerange{\pageref{firstpage}--\pageref{lastpage}}
\maketitle

\begin{abstract}
We present the detection of [$3.6~\mu$m] photometric variability in two young, L/T transition brown dwarfs, WISE J004701.06+680352.1 (W0047) and 2MASS J2244316+204343 (2M2244) using the \textit{Spitzer Space Telescope}. We find a period of $16.4\pm0.2~$hr and a peak-to-peak amplitude of $1.07\pm0.04\%$ for W0047, and a period of $11\pm2~$hr  and amplitude of $0.8\pm 0.2\%$ for 2M2244. This period is significantly longer than that measured previously during a shorter observation. {We additionally detect significant $J$-band variability in 2M2244 using the Wide-Field Camera on UKIRT. } We determine the radial and rotational velocities of both objects using Keck NIRSPEC data. We find a radial velocity of $-16.0_{-0.9}^{+0.8}~$km s$^{-1}$ for 2M2244, and confirm it as a bona fide member of the AB Doradus moving group.
We find rotational velocities of $v \sin i=9.8\pm0.3~$km s$^{-1}$and $14.3^{+1.4}_{-1.5}~$km s$^{-1}$ for W0047 and 2M2244, respectively. With inclination angles of $85 ^{+5\circ}_{-9}$ and $76 ^{+14\circ}_{-20}$, W0047 and 2M2244 are viewed roughly equator-on. Their remarkably similar colours, spectra and inclinations are consistent with the possibility that viewing angle may influence atmospheric appearance. We additionally present \textit{Spitzer} $[4.5~\mu\mathrm{m}]$ monitoring of the young, T5.5 object SDSS111010+011613 (SDSS1110) where we detect no variability. For periods $<18~$hr, we place an upper limit of $1.25\%$ on the peak-to-peak variability amplitude of SDSS1110.
\end{abstract}




\section{Introduction}\label{sec:intro}

\parskip = 0cm 
The growing number of young exoplanets that have been directly imaged in the infrared \citep{Marois2008, Lagrange2010, Macintosh2015} have revealed some unexpected results.
With comparable temperatures but lower masses, the young directly imaged planets were expected to share similar atmospheric properties to the well-studied population of brown dwarfs. However most young directly-imaged planets appear much redder in the near-IR than their higher mass field counterparts with similar {spectral types.}
Fortunately, young brown dwarfs may still provide an excellent analogue to directly-imaged planets, and we now have a significant population of young brown dwarfs with colours and magnitudes similar to directly-imaged exoplanets, {many of which have estimated masses in the planetary-mass regime }{(see the compilation of young, red M and L dwarfs made by \citet{Faherty2016}, \citet{Liu2016} and references therein)}. 
Three such objects are WISEP J004701.06+680352.1 (W0047), 2MASS J2244316+204343 (2M2244) and SDSS J111010+011613 (SDSS1110) {\citep{Gizis2012, Knapp2004, Gagne2015a}.}
W0047 and SDSS1110 are kinematically confirmed members of the 150 Myr old AB Doradus moving group \citep{Bell2015}. 2M2244 is assigned a membership probability of $99.6\%$ for the same group based on its proper motion and distance \citep{Gizis2015, Gagne2015a, Gagne2014b, Liu2016}, but a radial velocity measurement is necessary to confirm moving group membership. {We measure its radial velocity in this paper (Section \ref{sec:ABDor}) and confirm it as a member of the AB Doradus moving group.}
W0047 is classified as an L7 INT-G brown dwarf 
 and 2M2244 is classified as an L6 VL-G object \citep{Gizis2015, Allers2013}. 
 W0047 and 2M2244 are a particularly interesting pair of young, low-gravity objects, with $0.65 - 2.5~ \mu$m spectra that are remarkably similar \citep{Gizis2015}.
 There are no other free-floating L/T transition dwarfs known to be both coeval and spectrally similar {that are bright enough for detailed characterisation (though see \citet{Best2015} for more candidates)}. SDSS1110 is a T5.5 $10-12~M_{\mathrm{Jup}}$ \citep{Gagne2015a} object, and is one of very few young, age-calibrated T dwarfs known to date. \added{W0047, 2M2244 and SDSS1110 are the lowest mass confirmed members of the AB Doradus moving group \citep{Liu2016}, and can thus provide powerful insights into the atmospheres of the directly-imaged planets.}
 
A key probe of  brown dwarf atmospheres is time-resolved photometric monitoring, which is sensitive to the spatial distribution of surface inhomogeneities as {objects} rotate. 
Large-scale field brown dwarf surveys have revealed ubiquitous variability across the entire L-T spectral range \citep{Buenzli2014, Radigan2014, Wilson2014, Metchev2015a}. 
 Due to their lower gravity, young brown dwarfs exhibit  different atmospheric scale height and time scales than old field brown dwarfs \citep{Freytag2010,Marley2010,Marley2012}. Thus, studying their variability provides valuable information on atmospheric structure  in brown dwarfs and exoplanet atmospheres as a function of surface gravity.
Because of their more recent formation, young brown dwarfs and exoplanets have inflated radii compared to the field brown dwarfs. Hence, they are expected to rotate more slowly {than} their {older} counterparts due to conservation of angular momentum.
However, the planetary mass objects $\beta$ Pic b, PSO-318.5-22 and 2M1207b all have rotation periods of $6-11~$hr \citep{Snellen2014, Biller2015, Allers2016, Zhou2015}, similar to higher mass brown dwarfs \citep{ZapateroOsorio2006}.

To date, the observed variability has been interpreted as evidence for condensate clouds, which are required by the majority of brown dwarf and exoplanet models \citep{Marley2010, Morley2014}. {Magnetic phenomena, such as starspots, have also been suggested as a driver of photometric variability. While some L-T type brown dwarfs have been found to possess strong magnetic fields \citep{Pineda2016, Kao2015}, \citet{Miles-Paez2017a} report no correlation between magnetic activity and photometric variability in  a sample of L0-T8 brown dwarfs. }Recently, \citet{Tremblin2016} proposed cloud-free models, suggesting that the observed variability is due to differing CO abundances or temperature fluctuations. Further work is required to establish which scenario is appropriate for these objects.

W0047 and 2M2244 present the unique opportunity to explore the effects of both viewing angle and age on observed variability. 
For an equator-on object  (with an inclination
angle, $i \sim90^{\circ}$) we measure the full variability amplitude via
photometric monitoring. In contrast, lower variability amplitudes  are measured for objects that are close to pole-on \citep{Vos2017}. Determining the variability amplitude and inclination angle of each object allows us to disentangle the effects of viewing angle on the observed variability.

\citet{Metchev2015a} find evidence for higher variability amplitudes for young L3-L5.5 objects. This is unexpected because atmospheric models for young objects typically require very thick clouds \citep{Madhusudhan2011} and variability studies  have suggested that older objects with patchy coverage of thinner and thicker clouds tend to have the highest variability amplitudes \citep{Apai2013, Buenzli2015a}.
2M2244, W0047 and SDSS1110 provide three valuable data points to further explore this trend beyond the early L-type dwarfs.

Periodic variability has previously been detected in W0047 and 2M2244. \citet{Lew2016} report variability with a {peak-to-peak $J$-band  amplitude of } $8\%$  for W0047 during a $9~$hr observation, determining a period of $13.2\pm0.14~$hr. \citet{Morales-Calderon2006} obtained \textit{Spitzer} [$4.5~\mu m$] time-resolved photometry {of 2M2244} and report variability with a period of $4.6~$hr and a {peak-to-peak} amplitude of $8~$mmag during a $5.7~$hr observation. SDSS1110 has no previous variability detections in the literature.
We have obtained \textit{Spitzer} photometric monitoring for W0047, 2M2244 and SDSS1110 {and $J-$band monitoring of 2M2244 taken with WFCAM at UKIRT,} as well as high dispersion NIRSPEC spectra of W0047 and 2M2244. { The spectrum of W0047 was first presented by \citet{Gizis2015}, and we use the same dataset in this paper. }
The paper is organised as follows. In Section \ref{sec:spec} we discuss the analysis and results of our Keck NIRSPEC high resolution spectra of 2M2244 and W0047. In Section \ref{sec:spitzer} we present the lightcurves of our three targets 2M2244, W0047 and SDSS1110. In Section \ref{sec:inclination} we calculate the inclination angles of W0047 and 2M2244.

\begin{figure*}
	\centering  
\begin{subfigure}
	\centering
		\includegraphics[scale=0.24]{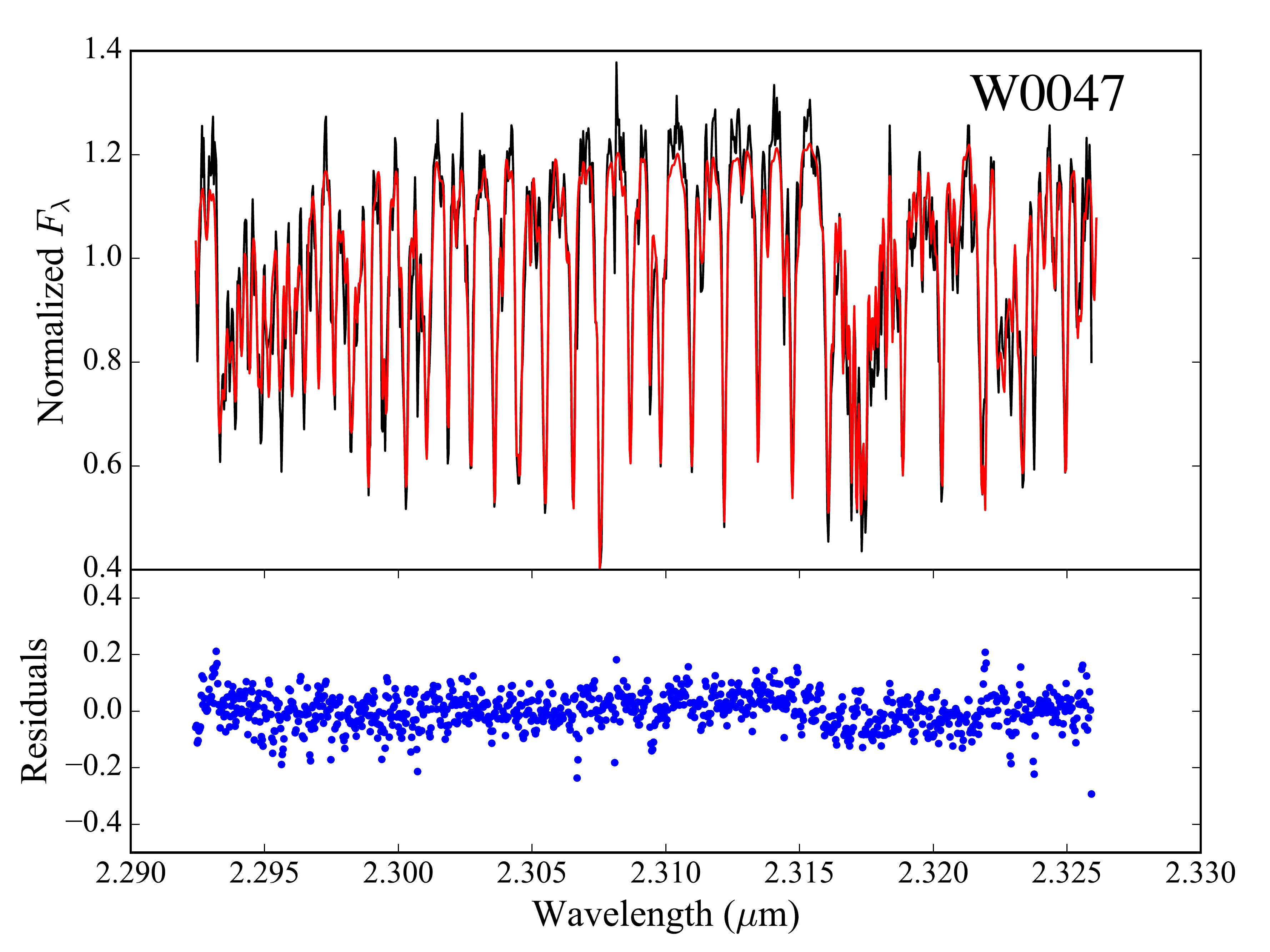}
\end{subfigure}
\begin{subfigure}
	\centering
		\includegraphics[scale=0.24]{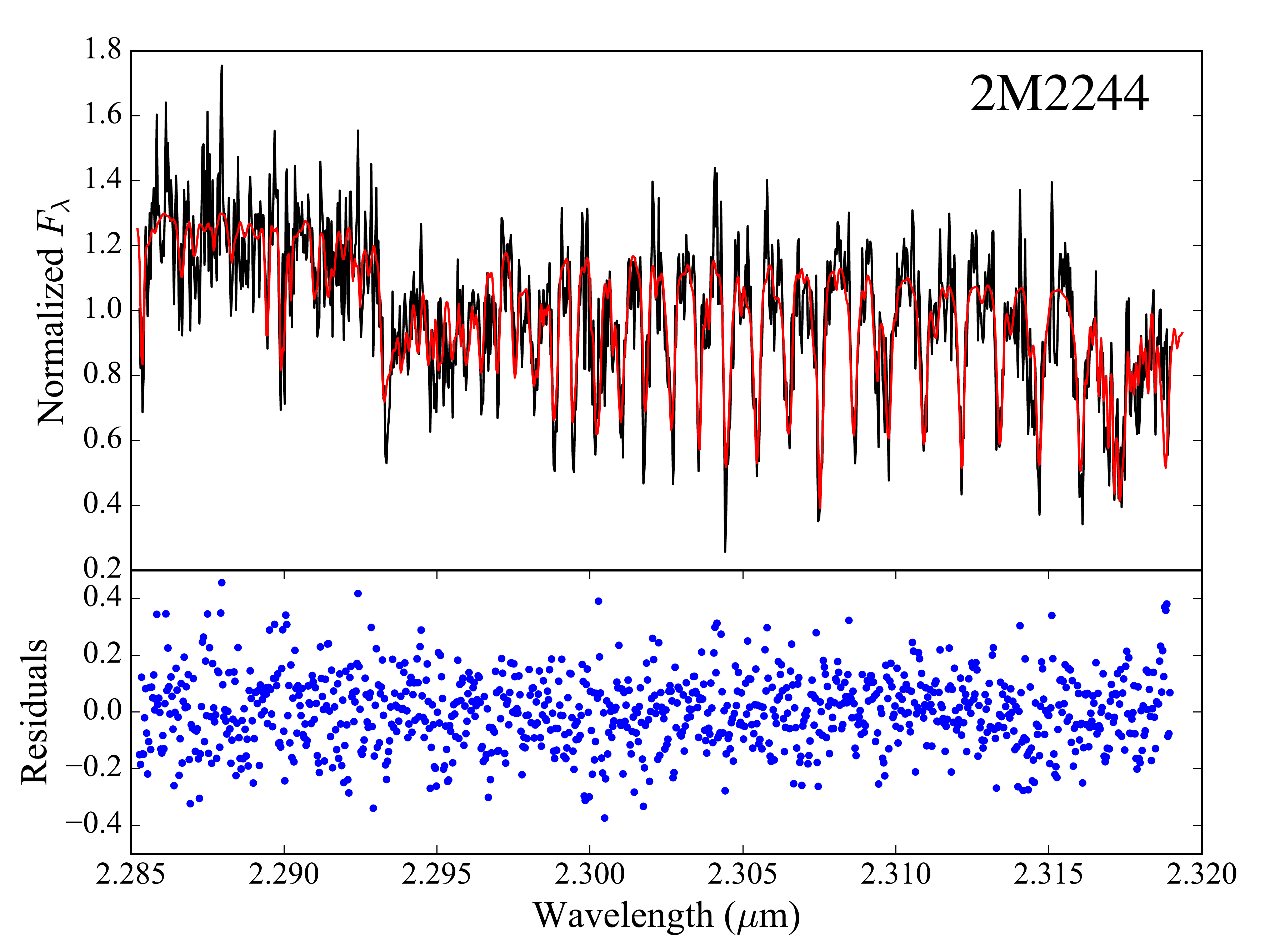}
\end{subfigure}
	\caption{{\it Left:} The observed spectrum of W0047 (black) compared to our best fit forward model (red). Residuals are plotted in the bottom panel. {\it Right:} The observed and best fit forward model of 2M2244.}
	\label{fig:spec}
\end{figure*}

\section{Keck NIRSPEC High Dispersion Spectroscopy}\label{sec:spec}

We obtained high dispersion NIRSPEC spectra for W0047 from the Keck Observatory Archive (Prog ID: U055NS, PI: Burgasser) and observed 2M2244 as part of a larger program (Prog ID: N160NS, PI: Allers). NIRSPEC is a near-infrared echelle spectrograph on the Keck II $10~m$ telescope on Mauna Kea, Hawaii. The NIRSPEC detector is a $1024 \times 1024$ pixel ALADDIN InSb array. Observations were carried out using the NIRSPEC-7 ($1.839 - 2.630~ \mu m$) passband in echelle mode using the 3 pixel slit ($0.432''$), echelle angles of $62^{\circ}.68-62^{\circ}.97$, and grating angles of $35^{\circ}.42-35^{\circ}.51$. Observations of targets were gathered in nod pairs, allowing for the removal  of sky emission lines through the subtraction of  consecutive images. Arc lamps were observed for wavelength calibration. $5-10$ flat field and dark images were taken for each target to account for variations in sensitivity and dark current on the detector.  {W0047 was observed on September 17 2013 with $2\times1200~$s exposures at an airmass of 1.5 and a mean DIMM seeing of $1.0\arcsec$. 2M2244 was observed on July 6 2013 with $4\times240~$s exposures at an airmass of 1.0 and a mean DIMM seeing of $0.5\arcsec$.}

We focus our analysis on order 33 ($2.286-2.326~\mu$m) since this part of the spectrum contains a good blend of sky lines and brown dwarf lines, allowing for an accurate fit. {This spectral region is rich in CO features, as well as H$_2$O and CH$_4$ features. These features are discussed in detail by \citet{Blake2010}.} Order 33 is also commonly used in the literature for NIRSPEC high dispersion N-7 spectra \citep{Blake2010, Gizis2013}. We additionally look at orders 32 {($2.364-2.398~\mu$m)} (for W0047) and 38 {($1.987-2.016~\mu$m)} (for both W0047 and 2M2244) to check for consistency.
Data were reduced using a modified version of the REDSPEC reduction package to spatially and spectrally rectify each exposure. The NIRSPEC Echelle Arc Lamp Tool was used to identify the wavelengths of lines in our arc lamp spectrum. 
After nod-subtracting pairs of exposures, we create a spatial profile which is the median intensity across all wavelengths at each position along the slit. 
We use Poisson statistics to determine the noise per pixel at each wavelength. We extract the flux within an aperture in each nod-subtracted image to produce two spectra of our source.  The extracted spectra are combined using a robust weighted mean with the xcombspec procedure from the SpeXtool package \citep{Cushing2004}.

\subsection{Determining Radial and Rotational Velocities}

We use the approach outlined in \citet{Allers2016} to determine the radial and rotational velocities of W0047 and 2M2244. We employ forward modelling to simultaneously fit the wavelength solution of our spectrum, the rotational and radial velocities, the scaling of telluric line depths, and the FWHM of the instrumental line spread function (LSF). We use the BT-Settl model atmospheres \citep{Allard2012} as the intrinsic spectrum for each of our targets. In total, the forward model has nine free parameters: the $T_\mathrm{eff}$ and $\log(g)$ of the atmosphere model, the $v_r$ and $v \sin i$ of the brown dwarf, $\tau$ for the telluric spectrum, the LSF FWHM, and the wavelengths of the first, middle and last pixels. The forward model is compared to our observed spectrum, and the parameters used to create the forward model are adjusted to achieve the best fit. 

To determine the best fit parameters of our forward model as well as their posterior distributions, we use a Markov Chain Monte Carlo (MCMC) approach. This involves creating forward models that allow for a continuous distribution  of $T_{\mathrm{eff}}$ and $\mathrm{log}(g)$ by linearly interpolating between atmosphere grid models. We employ the DREAM(ZS) algorithm \citep{terBraak2008}, which uses an adaptive stepper, updating model parameters based on chain histories. To ensure that the median absolute residual of the fit agrees with the median uncertainty of our spectrum, we include a systematic uncertainty of $1.4\%$ in the spectrum of W0047.We plot our spectra and best fit models along with the residuals in Figure  \ref{fig:spec}. 
Final  values for $v \sin i$ and radial velocities (RV) are shown in Table \ref{tab:results}. Their $1\sigma$ uncertainties are determined from their marginalised distributions obtained from our MCMC method. \added{Although we obtain values for $T_{\mathrm{eff}}$ and $\log (g)$, these derived values should not be considered physical since we are using a narrow wavelength range in K-band. Furthermore, atmospheric models are known to be unreliable for young L/T transition objects, even if J-band data are included \citep{Liu2013a,Allers2016}. These parameters are more reliably determined from evolutionary models, as is done in Section \ref{sec:props}.} The results for both 2M2244 and W0047 are consistent across orders 32, 33 and 38 {at the $2\sigma$ level.} { The mean and standard deviation of the LSF FWHM is $0.08\pm0.01~$nm and $0.081\pm0.002~$nm for 2M2244 and W0047 respectively, resulting in a resolution $R=\lambda /\Delta  \lambda \simeq 29000$ for both objects. The precision of our wavelength solution is determined to be $0.0025~$nm and $0.0124~$nm for 2M2244 and W0047.}

Our $v\sin i$ measurement of $9.8\pm0.3~$km s$^{-1}$for W0047 is higher than both previous measurements by \citet{Gizis2015} ($4.3\pm2.2~$kms$^{-1}$) and \citet{Lew2016} ($6.7^{+0.7}_{-1.4}~k$ms$^{-1}$), despite all three measurements using the same dataset. The model atmosphere for W0047 used by \citet{Gizis2015} has $T_{\mathrm{eff}}=2300$ and $\log(g)=5.5$ while evolutionary models predict $T_{\mathrm{eff}}=1270$ and $\log(g)=4.5 $ \citep{Gizis2015}. Our model (with   $T_{\mathrm{eff}}=1670$ and $\log(g)=5.2$) is in better agreement with the evolutionary model. Higher effective temperature and surface gravity results in more pressure broadening, producing a lower value of $v\sin i$. \citet{Lew2016} do not provide details on the atmospheric model used. Again, the consistency betweens orders 32, 33 and 38 further supports our results.

\subsection{2M2244+20 Membership in AB Doradus}\label{sec:ABDor}
A radial velocity measurement is required to confirm moving group membership.
Using Bayesian analysis to assess the membership of $>$M$5$ brown dwarfs, \citet{Gagne2014c} find a $99.6\%$ probability that 2M2244 is a member of the AB Doradus moving group, predicting a radial velocity of $-15.5\pm1.7~$km s$^{-1}$. Our measured radial velocity of $-16.0\pm0.9~$km s$^{-1}$ is consistent with the predicted radial velocity.
Including the measured radial velocity, along with parallax and proper motion measurements from \citet{Liu2016}, and using the BANYAN-II web tool \citep{Gagne2014c, Malo2013}, 
 the probability of AB Doradus membership increases to $99.96\%$. Thus, our radial velocity measurement confirms 2M2244 as a member of the AB Doradus moving group.
 
 \subsection{The Physical Properties of W0047 and 2M2244}\label{sec:props}
\added{ \citet{Filippazzo2015} provide radius, $\log (g)$, $T_{\mathrm{eff}}$ and mass estimates from evolutionary models for W0047 and 2M2244, however, the estimated age range used in this analysis of $50-110~$Myr for AB Dor is systematically younger than current estimates. \citet{Barenfeld13} place a strong lower limit of $110~$Myr and \citet{Luhman2005} provides an upper limit of $150~$Myr on the age of AB Dor. Furthermore, \citet{Filippazzo2015} use a kinematic distance to determine the luminosity of 2M2244 while \citet{Liu2016} has since measured its parallax. 
We use these measured parallaxes to update the luminosities of 2M2244 and W0047. The errors on the updated luminosities are slightly overestimated, since the bolometric magnitudes and errors are not given in \citet{Filippazzo2015}.
For a uniformly-distributed age of $110-150~$Myr and normally-distributed luminosities, we determine the physical properties of 2M2244 and W0047 using model isochrones (final parameters shown in Table \ref{tab:evo}). W0047 and 2M2244 both exhibit extremely red $J-K$ colours, indicating a dusty atmosphere. Thus, we use the \citet{Saumon2008} solar metallicity $f_{sed}=2~$  models.}
\added{The older age of the AB Doradus moving group that is used in this analysis pushes both masses above the deuterium burning limit, and above the masses presented in  \citet{Filippazzo2015}. The revised radii are consistent with those reported by \citet{Filippazzo2015}. }

\begin{table}
\centering
\caption{\added{Physical properties of W0047 and 2M2244 from the \citet{Saumon2008} $f_{sed}=2~$ evolutionary model.}}
\label{tab:evo}
\begin{tabular*}{\columnwidth}{l@{\extracolsep{\fill}}ll}
\hline \hline
 	           				 & W0047                & 2M2244                 \\[4pt]
				 \hline
$\log (L/L_{\circ})$        		& $-4.44 \pm 0.04$         		& $-4.48 \pm 0.02$     \\[2pt]
Mass ($M_{Jup}$)   			& $19.5^{+1.6}_{-1.7}$ & $19.0^{+1.4}_{-1.5}$   \\[2pt]
$T_{\mathrm{eff}}$ 			& $1250^{+20}_{-30}$   & $1230^{+16}_{-15}$     \\[2pt]
Radius ($R_{Jup}$) 			& $1.28 \pm 0.02$       & $1.28 \pm 0.02$         \\[2pt]
$\log (g)$  (dex)     			& $4.49 \pm 0.05$      & $4.48^{+0.04}_{-0.05}$\\[2pt]
\hline 
\end{tabular*}
\end{table}

\section{Spitzer and WFCAM Photometry}\label{sec:spitzer}

 \begin{figure}
	\centering
		\includegraphics[scale=0.42]{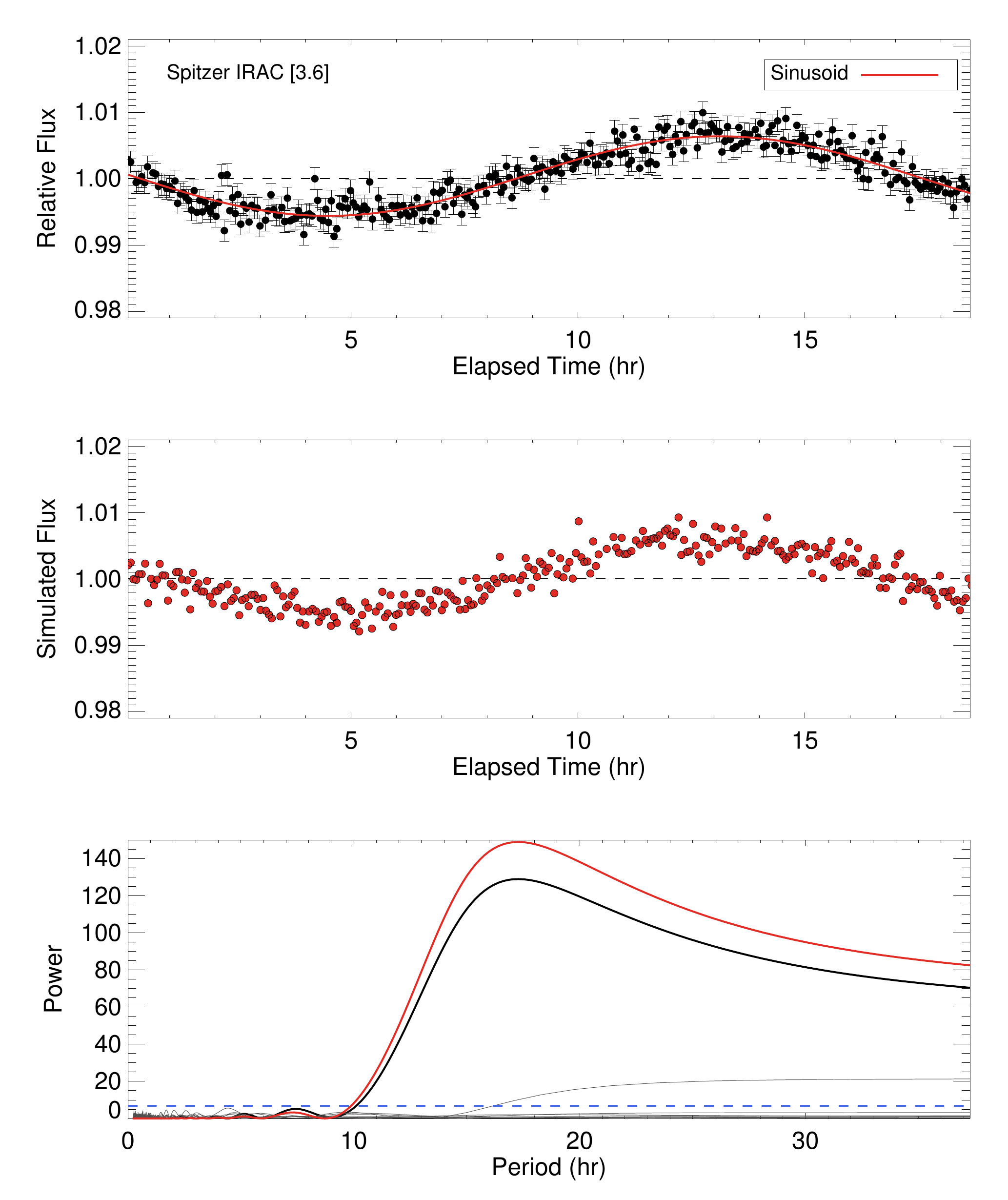}
	\caption{Top panel shows the normalised, pixel phase corrected lightcurve of W0047 with best-fit sinusoidal function overplotted in {red}. The best-fit function gives a period of $16.3 \pm 0.2~$hr and an amplitude of $1.08\pm 0.04 \%$. The middle panel shows the best fit function injected into a simulated lightcurve. The bottom panel {shows} the periodogram of the target and the simulated curve, as well as the periodogram of several reference stars in the field. The blue dashed line shows the $1\%$ false-alarm probability.}
	\label{fig:lcurveW0047}
\end{figure}

For our \textit{Spitzer} observations of W0047, 2M2244 and SDSS1110 we followed standard observing practices for obtaining precise, stable, and nearly-photon limited performance. 
 We employed 'staring mode' AORs in which the object did not move on the chip throughout the entire observation, with a long exposure time \citep{Metchev2015a}. {
W0047 and 2M2244 were observed for $18.7~$hr and $8.8~$hr on January 9 and September 15 2016 respectively, in the \textit{Spitzer} [$3.6~\mu $m] band with an exposure time of $30~$s and a pixel scale of $1.221\arcsec$. SDSS1110 was observed for $9.0~$hr on April 5 2016 in the \textit{Spitzer} [$4.5~\mu $m] band with an exposure time of $100~$s and a pixel scale of $1.231\arcsec$.} Additionally, we include \textit{Spitzer} [$4.5~\mu $m] archival data of 2M2244 (Program ID: 20079, PI: Stauffer) and published in \citet{Morales-Calderon2006} for re-analysis.

\begin{figure*}
	\centering
		\includegraphics[scale=0.55]{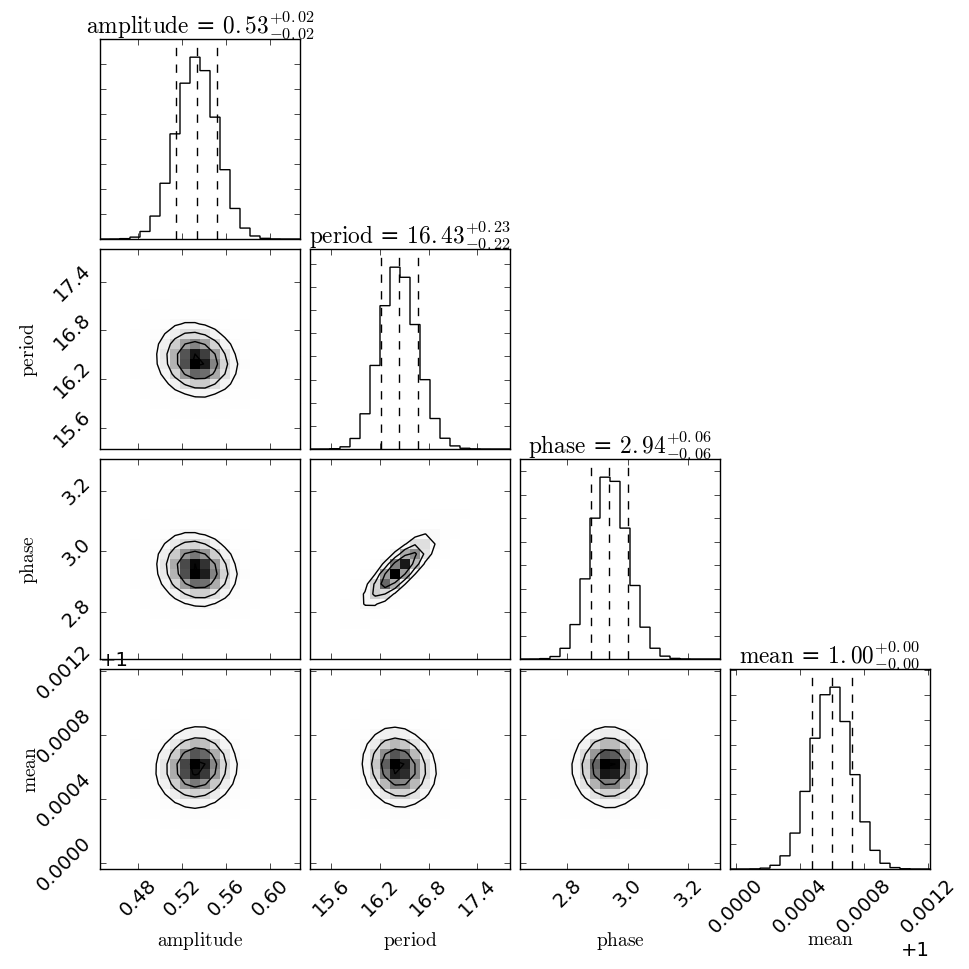}
	\caption{{Posterior distributions of parameters of the \textit{Spitzer} lightcurve of W0047 (shown in Figure \ref{fig:lcurveW0047}) The middle dashed line is the median, the two outer vertical dashed lines represent the $68\%$ confidence interval. The contours show the $1\sigma$, $1.5\sigma$ and $2\sigma$ levels.}}
	\label{fig:tri_W0047}
\end{figure*}

Photometry was obtained from the corrected Basic Calibrated Data images, provided by the Spitzer Science Center after processing through IRAC pipeline version 19.2.0. The centroids of the target and a number of reference stars are found using \textit{box\_centroider.pro}. We perform aperture photometry about these centroids, using various aperture sizes and choosing the aperture size that produces lightcurves with the lowest RMS (3.0, 3.5 and 3.5 pixels for {2M2244}, {W0047} and SDSS1110 respectively).
{Outliers are identified and rejected from the raw light curves using a $6\sigma$ clip, {removing $\sim5 - 45$ points in each lightcurve.}}

The light curves are then corrected for intrapixel sensitivity variations, the so-called 'pixel phase effect'. This is the slight variation in flux depending on where a point source falls with respect to the centre of a pixel. The pixel phase response is modelled as a `double-gaussian', a summation of gaussians in the orthogonal pixel directions. We correct for the pixel phase effect using the \textit{pixel\_phase\_correct\_gauss.pro} routine from the \textit{Spitzer} IRAC website. The pixel phase corrected flux is then binned into $\sim5$ minute bins {using a weighted average}, followed by a final $3\sigma$ clip to produce the final lightcurves. 
{The photometric noise of our normalised and corrected light curves is calculated following \citet{Radigan2014}. While the standard deviation produces a measurement of noise for flat curves, in the case of variable lightcurves the standard deviation measures both noise and intrinsic variations. We therefore use the point-to-point noise to measure the photometric noise. This is the standard deviation of the lightcurve subtracted from a shifted version of itself, divided by $\sqrt 2$. This measure of photometric noise is not sensitive to low frequency trends and thus provides a better estimate of the noise for variable lightcurves.}

{ We also include an observation of 2M2244 taken with the infrared Wide-Field Camera (WFCAM; \citet{Casali2007}) on July 21 2016 UT as part of a larger survey for variability on free-floating low-mass objects. This is a wide-field imager on the $3.8~$m UK Infrared Telescope on Mauna Kea, with a pixel scale of 0.4''. The observation was carried out with the $J$-band filter and seeing of $\sim1.1"$ during the $4~$hr sequence. The target was observed using an ABBA nod pattern, with 3 exposures of $40~$s at each position. The frames were reduced using the WFCAM reduction pipeline \citep{Irwin2004, Hodgkin2009} by the Cambridge Astronomical Survey Unit. Aperture photometry is performed on the target as well as a large number of reference stars in the field using an aperture size of 3.5 pixels. Raw light curves obtained from aperture photometry display brightness fluctuations due to changes in seeing, airmass and residual instrumental effects. To a very good approximation
these changes are common to all stars in the field of view and can be removed via division of a 
calibration curve calculated from a set of iteratively chosen, well-behaved reference stars \citep{Radigan2012}. For each star a calibration curve is created by median combining all other reference stars (excluding that of the target and of the star itself).
The standard deviation and linear slope for each light curve is calculated and stars with a standard deviation or slope $1.2$ times 
greater than that of the target are discarded. This process is iterated a number of times, until a set of 
well-behaved reference stars is chosen. Reference stars are also examined by eye to check for any residual trends. Final detrended lightcurves are obtained by dividing the raw curve for each
star by its calibration curve. }

\subsection{Identification of Variables}

We plot the periodogram of the target as well as a number of reference stars in the field to identify periodic variability in our targets. For each periodogram, the $1\%$ false-alarm probability (FAP) is calculated from 1000 simulated lightcurves. These lightcurves are produced by randomly permuting the indices of reference star lightcurves \citep{Radigan2014}. This produces lightcurves with Gaussian-distributed noise. The $1\%$ FAP is plotted in blue in each periodogram. The rotational periods and peak-to-peak variability amplitudes of targets showing periodic variability are determined by fitting an appropriate function to the data using \textit{mpfit.pro}. This is an implementation of the Levenberg-Marquardt least-squares minimisation algorithm which provides the best-fit periods and variability amplitudes with their $1\sigma$ uncertainties. Finding that the least-squares method can be sensitive to initial parameter guesses, we also use the MCMC algorithm \textit{emcee} \citep{fm2013} to fully explore the posterior probability distributions of our model parameters.

Aperiodic or stochastic variations are not easily detectable from Lomb-Scargle periodograms so we additionally check for stochastic variability by comparing the photometric standard deviation of our target with the mean standard deviation of comparison stars of similar brightness. If the standard deviation of the target is considerably larger than the mean standard deviation of the comparison stars this suggests stochastic variability in the target.

\subsection{W0047}
The lightcurve of W0047 (Figure \ref{fig:lcurveW0047}) appears sinusoidal over an entire period. The periodogram displays a strong peak at $\sim16~$hr that is well above the $1\%$ FAP value.
The least-squares best-fit sinusoidal function gives a period of $16.3 \pm 0.3~$hr and an amplitude of $1.08\pm 0.04 \%$. We also use the \textit{emcee} package \citep{fm2013} to obtain the full posterior probability distribution for each parameter of the sinusoid model. We use 1000 walkers with 7500 steps (after discarding the initial burn-in sample) in the four-dimensional parameter space to model the lightcurve.
 Figure \ref{fig:tri_W0047} shows the posterior probability distributions of the amplitude, period, phase and constant parameters of the fit. Each parameter is well constrained, and the MCMC method gives a period of $16.4\pm0.2~$hr and a peak-to-peak amplitude of $1.07\pm0.04\%$.

Assuming rigid rotation, we use our measured $v\sin i$ and a  radius estimate of {$1.28\pm0.02~R_{\mathrm{Jup}}$} allow us to place an upper limit of \added{$16.3^{+0.8}_{-1.4}~$hr} on the rotational period of W0047. We can therefore discount the possibility of a double-peaked lightcurve with a longer rotational period.
The measured period is significantly longer than the previously measured $13.2\pm0.14~$hr  \citep{Lew2016}, however this initial period was determined from a $\sim 9~$hr observation that did not cover a full rotation.
{The photometric noise of our target is similar to that measured for comparison stars in the field of similar brightness; thus we find no evidence for aperiodic variability.}

With a peak-to-peak amplitude of $1.07\pm0.04\%$, this is among the highest \textit{Spitzer} [$3.6~\mu$m] variability amplitudes detected. \citet{Metchev2015a} notes a tentative correlation between low-gravity and high amplitude variability among a sample of eight L3-L5.5 dwarfs. The variability detection measured here adds to a growing number of young, L objects that display high amplitude variability, suggesting that this correlation may extend into the late-L spectral types \citep{Metchev2015a, Biller2015, Lew2016}.

\begin{figure}
	\centering
		\includegraphics[scale=0.42]{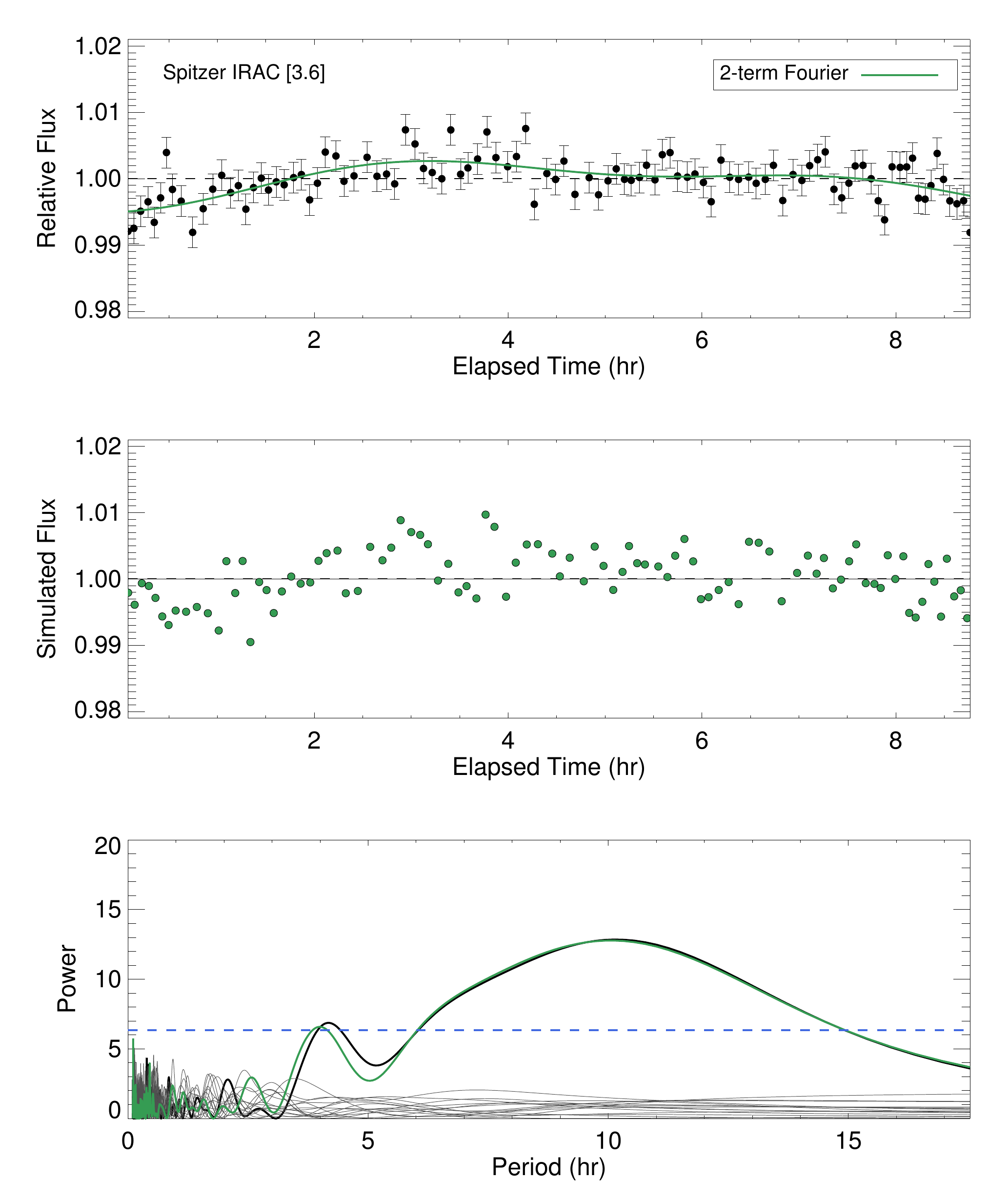}
	\caption{Same as Figure \ref{fig:lcurveW0047}, but for 2M2244 data taken on Sep 15 2016.  Here we consider a two-term Fourier series to model the variability. The Fourier function gives a period of $10 \pm 2.4~hr$ with a peak to trough amplitude of $0.8\pm0.2\%$. The bottom panel shows that the Fourier-term fit matches the target periodogram well, reproducing both the minor smaller peak at $\sim4~$hr and the large peak at $\sim10~$hr.}
	\label{fig:lcurve2M2244}
\end{figure}

\begin{figure*}
	\centering
		\includegraphics[scale=0.5]{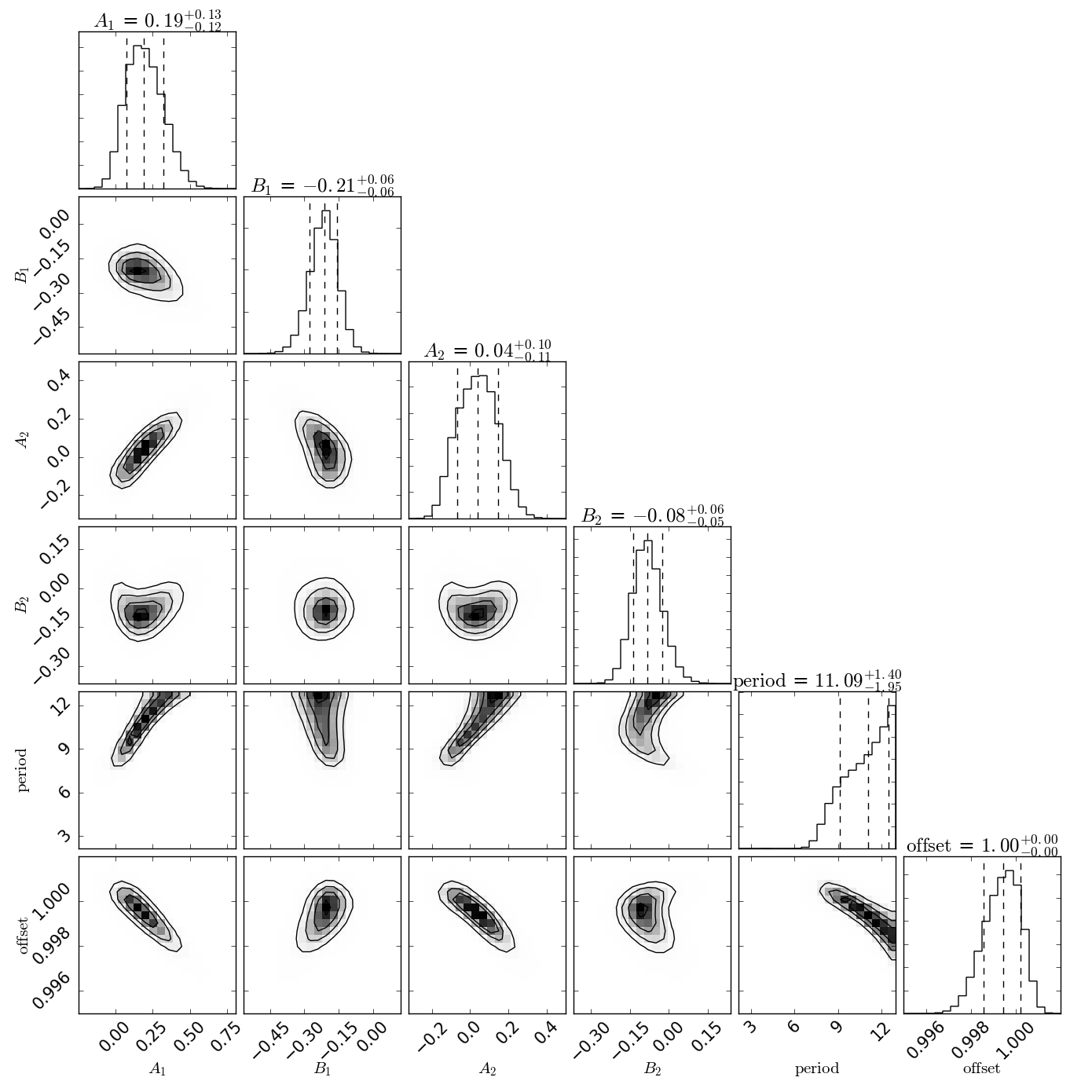}
	\caption{{Posterior distributions of parameters of the Fourier model fit to the \textit{Spitzer} lightcurve of 2M2244 (shown in Figure \ref{fig:lcurve2M2244}). The middle dashed line is the median, the two outer vertical dashed lines represent the $68\%$ confidence interval. We have placed an upper limit on the period of  $13~$hr using our radius estimate from Table \ref{tab:evo} and $v \sin i$ measurement from Table \ref{tab:results} . The contours show the $1\sigma$, $1.5\sigma$ and $2\sigma$ levels.}}
	\label{fig:tri_W2244_2}
\end{figure*}

\subsection{2M2244}

\subsubsection{Spitzer [$3.6~\mu$m] Monitoring}
In contrast to W0047, the \textit{Spitzer} [$3.6~\mu$m] lightcurve of 2M2244 does not appear sinusoidal (Figure \ref{fig:lcurve2M2244}).  {The photometric noise of 2M2244 is similar to the noise measured in  comparison stars of similar brightness in the field. Thus we do not detect any stochastic or aperiodic variability for 2M2244. } \citet{Morales-Calderon2006} report a sinusoidal light curve period of $4.6~$hr for this object; however the latest observations look very different. The periodogram shows a small peak at $\sim4~$hr that is approximately at the $1\%$ FAP level which{ roughly coincides with the $4.6~$hr period determined by \citet{Morales-Calderon2006}. } We also identify a broad peak at $\sim 9.6~$hr that is highly significant.
 {The light curve does not exhibit a sinusoidal shape,} so we consider a two-term truncated Fourier series, which is an appropriate model for more complex lightcurves \citep{Heinze2014, Yang2016}. {This model describes a scenario in which two atmospheric features are located on either hemisphere of the brown dwarf, each causing changes in brightness as they rotate in and out of view. }The two-term Fourier series is given by:
\begin{equation}
\centering
	F(t) = a_0 + \sum_{n=1}^{2} A_i \mathrm{sin} \left( \frac{2 \pi t}{P/i} \right) + B_i \mathrm{cos} \left( \frac{2 \pi t}{P/i} \right)
\end{equation}
The least-squares fit requires a `first guess' for the parameters, which we set to the peak of the periodogram for the period and 1 for all other parameters.  The least-squares best fit Fourier series model gives a period of $10.0\pm2.4~$hr. We inject this function into simulated lightcurves and reference stars to compute their periodograms. As seen in the bottom panel of Figure \ref{fig:lcurve2M2244}, the two-term Fourier signal produces a periodogram shape very similar to that of 2M2244, with a strong peak at $\sim10~$hr and a smaller peak at $\sim4~$hr.

After experimentation with different starting parameters for the least-squares fit, we find that the results are not consistent across different initial guesses for the model parameters. Using the \citet{Morales-Calderon2006} measurement of $4.6~$hr as an initial guess on the period of 2M2244,
 the best-fit solution gives a period of $\sim4~$hr. In contrast, using the peak of our periodogram ($\sim10~$hr) as an initial guess on the period, we obtain a best-fit period of $10~$hr for the Fourier model. In fact, any initial guess $>5~$hr yields a best-fit period of  $10~$hr. It is clear that the least-squares fitting procedure cannot locate global minima, and is over-dependent on initial guesses. Hence, we  use the \textit{emcee} algorithm to explore the posterior distribution of the model parameters using the two-term Fourier model. We use 1000 walkers with 7500 steps (after discarding the initial burn-in sample) to model the lightcurve. 
Our measured $v\sin i$ value of $14.3^{+1.3}_{-1.5}$ km s$^{-1}$ and estimated radius of $1.28\pm0.02~R_{\mathrm{Jup}}$ allow us to place an upper limit of $11.1^{+1.9}_{-1.2}~$hr on the period of 2M2244, hence we use an upper limit of $13~$hr as a prior in our MCMC analysis.
The posterior distributions of the parameters 
for the Fourier model are shown in Figure \ref{fig:tri_W2244_2}. 
This model favours a period of $11.1^{+1.4}_{-2.0}$ hr and this value is insensitive to the initial parameter guesses.

\begin{figure}
	\centering
		\includegraphics[scale=0.42]{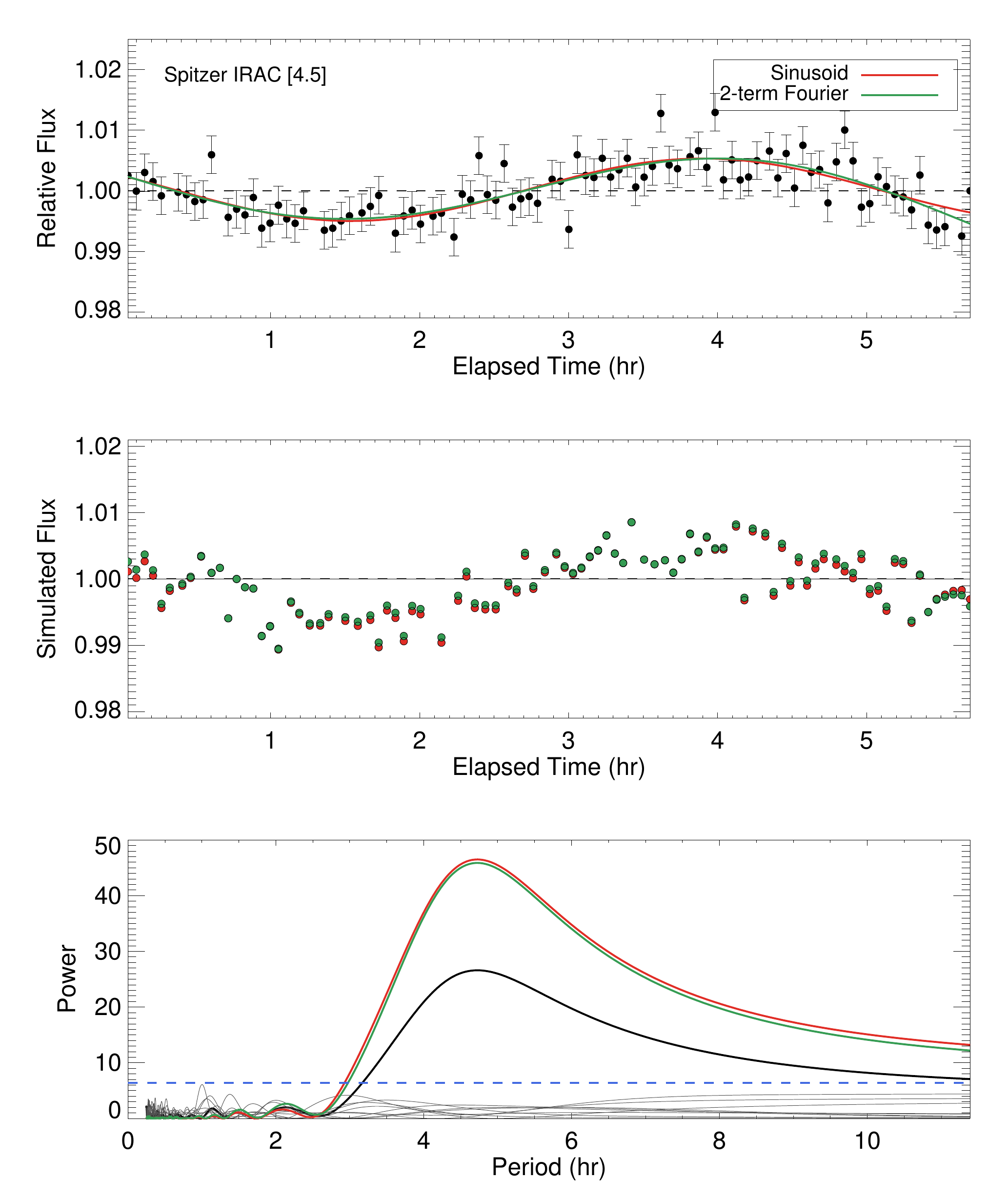}
	\caption{Same as Figure \ref{fig:lcurveW0047}, but for the \citet{Morales-Calderon2006} [$4.5~\mu m$] observation of 2M2244, taken on November 27 2005. The best-fit sinusoid function gives a period of $4.6 \pm 0.2~$hr while the  double-peaked Fourier function gives a period of $10\pm3~$hr. Injecting both functions into simulated lightcurves and reference stars gives a periodogram shape similar to the observed lightcurve's periodogram. The functions are indistinguishable from each other over this observation. }
	\label{fig:lcurve2M2244_MC}
\end{figure}

\begin{figure*}
	\centering
		\includegraphics[scale=0.55]{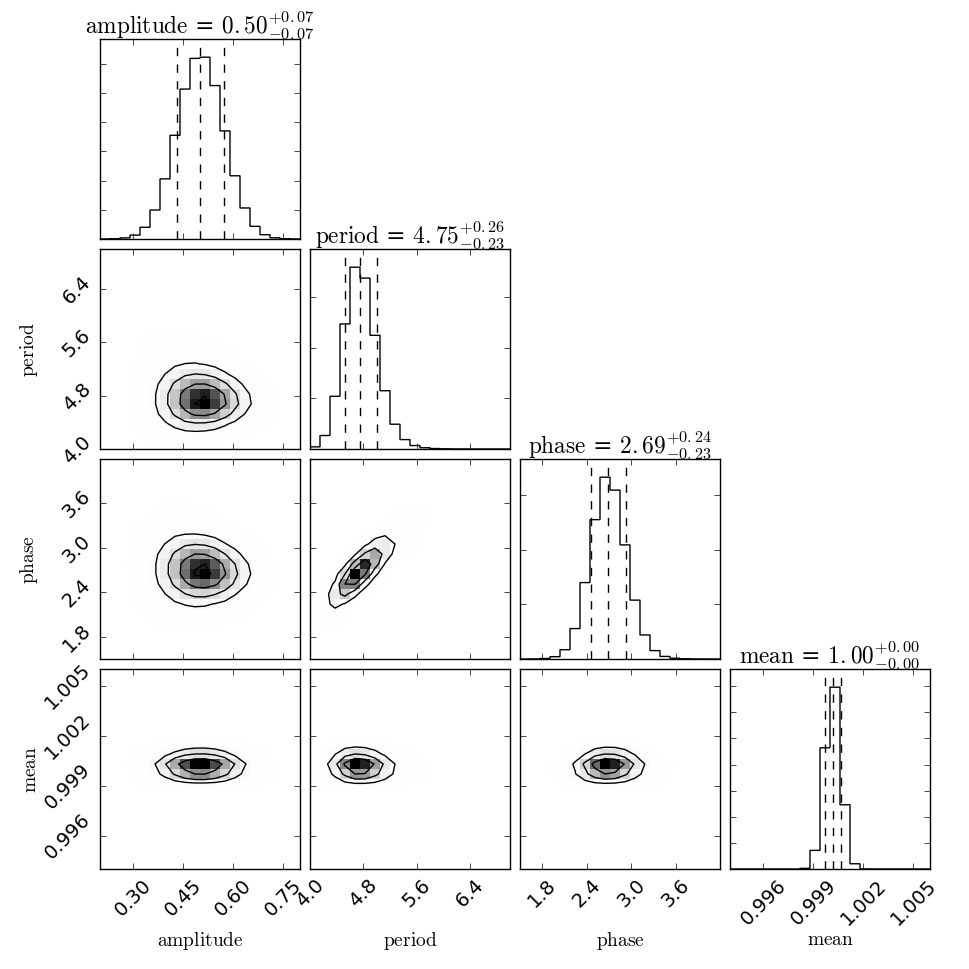}
	\caption{{Posterior distributions of parameters of the sinusoid fit to the  \citet{Morales-Calderon2006} \textit{Spitzer} lightcurve of 2M2244 (shown in Figure \ref{fig:lcurve2M2244_MC}) The middle dashed line is the median, the two outer vertical dashed lines represent the $68\%$ confidence interval. We have placed an upper limit on the period of  $13~$hr using our radius estimate from Table \ref{tab:evo} and $v \sin i$ measurement from Table \ref{tab:results}. The contours show the $1\sigma$, $1.5\sigma$ and $2\sigma$ levels.}}
	\label{fig:tri_MC_sin}
\end{figure*}

\begin{figure*}
	\centering
		\includegraphics[scale=0.5]{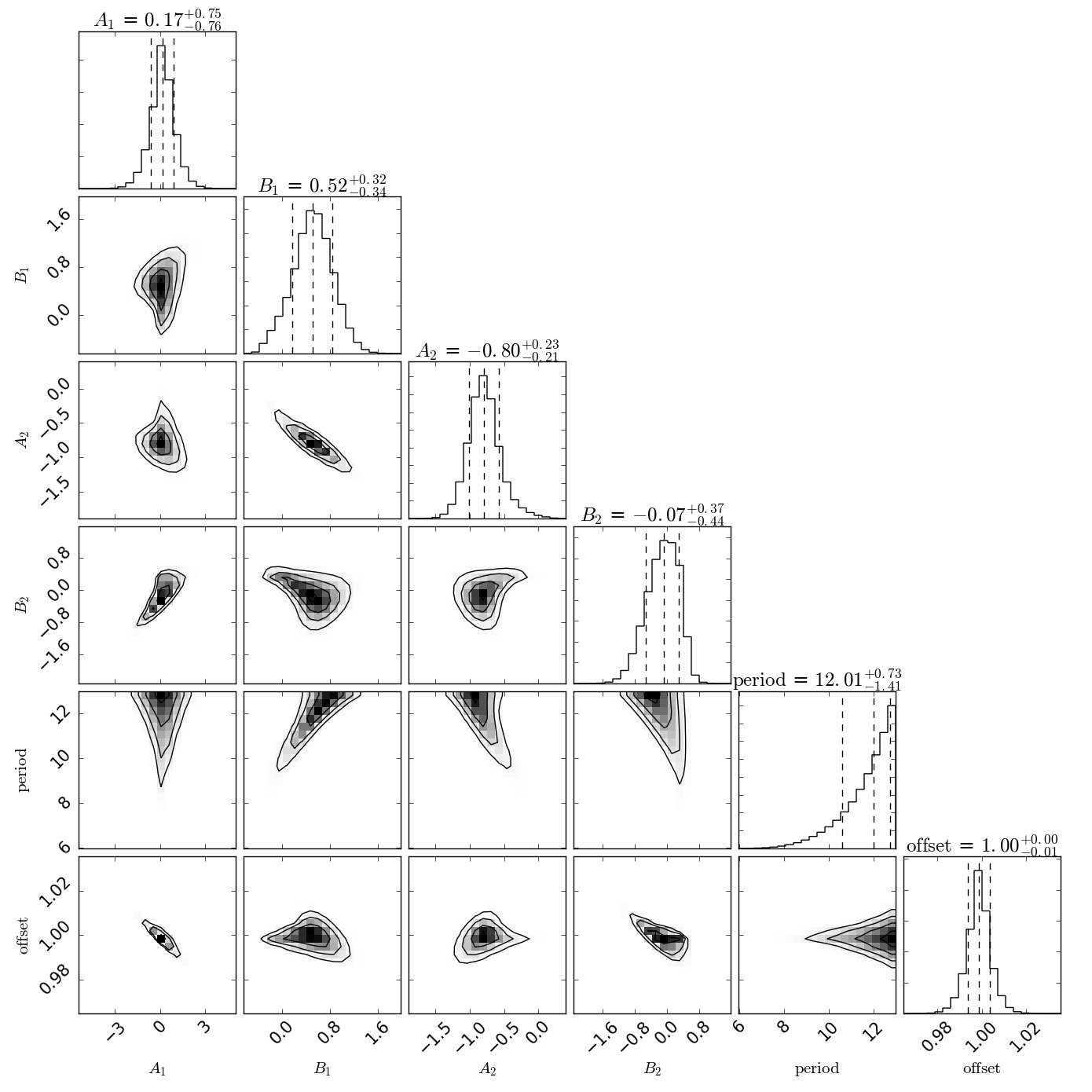}
	\caption{{Posterior distributions of parameters of the Fourier model fit to the  \citet{Morales-Calderon2006} \textit{Spitzer} lightcurve of 2M2244 (shown in Figure \ref{fig:lcurve2M2244_MC}) The middle dashed line is the median, the two outer vertical dashed lines represent the $68\%$ confidence interval. We have placed an upper limit on the period of  $13~$hr using our radius estimate from Table \ref{tab:evo} and $v \sin i$ measurement from Table \ref{tab:results}. The contours show the $1\sigma$, $1.5\sigma$ and $2\sigma$ levels.}}
	\label{fig:tri_MC_Fourier}
\end{figure*}

\subsubsection{Spitzer [$4.5~\mu$m] Monitoring}

Our measured period is inconsistent with that of \citet{Morales-Calderon2006} who find a period of $4.6~$hr during a $\sim6~$hr observation in the \textit{Spitzer} [$4.5~\mu$m] band. We downloaded these data from the Spitzer Heritage Archive. The reduced lightcurve and periodogram are shown in Figure \ref{fig:lcurve2M2244_MC}. The periodogram peaks at $4.6~$hr, as reported by \citet{Morales-Calderon2006}. The curve appears sinusoidal over the observation period but we investigate the possibility of a double-peaked lightcurve. Fitting a pure sinusoid to the data gives a period of $4.6\pm0.2~$hr
while fitting a two-term truncated Fourier series gives a period of $10\pm3~$hr; however the functions are indistinguishable from each other over this observation. Injecting the $4.6~$hr sinusoid fit and the $10~$hr truncated Fourier fit into simulated lightcurves and reference stars produces the same periodogram shape as the target, seen in the bottom panel of Figure \ref{fig:lcurve2M2244_MC}. {We use the MCMC method to explore the parameter posterior distributions for both the sinusoid model and the Fourier model. Again we use an upper limit of $13~$hr as a prior on the period. The posteriors are shown in Figures \ref{fig:tri_MC_sin} and \ref{fig:tri_MC_Fourier}. Again, both models fit the light curve well, with the sinusoidal model giving a period of $4.8^{+0.3}_{-0 .2}~$hr and the Fourier series model giving a period of $12.01^{+0.7}_{-1.4}~$hr.} Thus, we conclude that the original   observation is too short to rule out a double-peaked lightcurve with the $\sim11~$hr period of the \textit{Spitzer} [$3.6~\mu$m] dataset, and from this dataset either scenario is possible.

\subsubsection{UKIRT WFCAM Monitoring}
\begin{figure}
	\centering
		\includegraphics[scale=0.42]{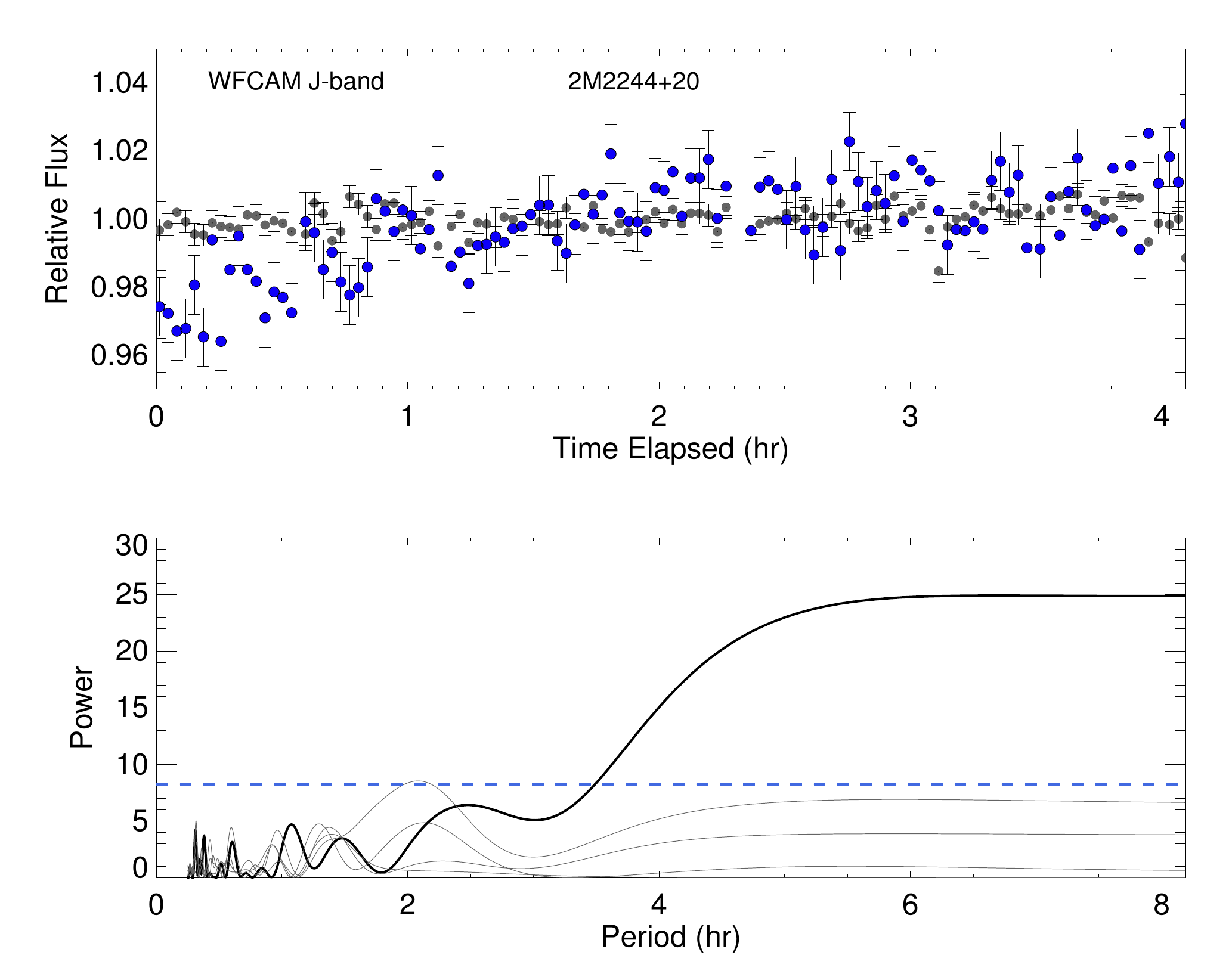}
	\caption{ {Upper panel: UKIRT WFCAM photometry of 2M2244+20 taken on 2016-06-21. 2M2244+20 is shown in blue with a reference star shown in grey. Lower panel: Periodogram of 2M2244+20 as well as the periodograms of reference stars in the field. In this observation 2M2244+20 shows trends with periodicities $<5.5~$hr.}}
	\label{fig:lcurveWFCAM2244}
\end{figure}

{The WFCAM photometry of 2M2244 is shown in Figure \ref{fig:lcurveWFCAM2244}. In this $4~$hr $J$-band observation we see evidence of significant ($\sim4\%$) variability. The periodogram shows a highly significant peak that favours periodicities $>5.5~$hr. This observation is too short to accurately measure the period, but it is consistent with a $\sim 11~$hr period. Since we have not covered a full period we cannot measure the full $J-$band variability amplitude, but can set a lower limit of $\sim4\%$.}

\subsubsection{The Period of 2M2244}
{Considering all three epochs of data for 2M2244, we favour a longer period of $11\pm2~$hr. We conclude that the initial \textit{Spitzer} [$4.5~\mu$m]  monitoring observation by \citet{Morales-Calderon2006} is too short to completely rule out a longer period. The light curve is most likely double-peaked in this epoch, due to two different atmospheric structures in either hemisphere. We see a very different shape in the September 2016 \textit{Spitzer} [$3.6~\mu$m] light curve. This is plausibly due to evolution of the cloud structure in the $\sim10$ years between epochs. This could also be due to the fact that we are probing different pressure levels in each \textit{Spitzer} band, however recent studies have found [$3.6~\mu$m]  and [$4.5~\mu$m] light curves to have similar shape and phase \citep{Metchev2015a, Cushing2016}. \added{A recent paper by \citet{Apai2017} suggests another possible explanation for evolving lightcurves such as that observed for 2M2244. In this paper the variability of three brown dwarfs is modeled by longitudinal bands with sinusoidal surface brightness modulations and an elliptical spot. When two bands have slightly different periods due to differing velocities or directions, they interfere to produce beat patterns. These beat patterns produce high amplitude variability when the waves are in phase and produce double-peaked variability when the phase shift between the waves is close to $90^{\circ}$. This model can explain light curves that are sometimes single-peaked and other times double-peaked as well as providing an explanation for the shape of the periodogram in the bottom panel of Figure \ref{fig:lcurve2M2244}, where the higher frequency peak at $\sim 5~$hr may be explained by a beat pattern with wavenumber $k=2$.}
 
 Both our periodogram and MCMC analysis of the new [$3.6~\mu$m] dataset point to a period of $\sim11.0~$hr. This period is also consistent with our UKIRT WFCAM $J-$band observation. As we still have not covered a full period for 2M2244 we combine the periods obtained from our  MCMC  Fourier models (shown in Figure \ref{fig:tri_W2244_2} and \ref{fig:tri_MC_Fourier}) to make a conservative estimate of $11\pm2~$hr for 2M2244.}
 
{ The observed peak-to-peak amplitude of $0.8\pm0.2\%$ for the more recent \textit{Spitzer} [$3.6~\mu$m]  is comparable to the $1.0\pm0.1\%$ modulation observed in the original \textit{Spitzer} [$4.5~\mu$m] epoch of \citet{Morales-Calderon2006}. The amplitude ratio, $A[4.5]/A[3.6]$ of $1.25\pm0.2\%$ is similar to the amplitude ratios found by \citet{Metchev2015a}. }

\begin{figure}
	\centering
		\includegraphics[scale=0.42]{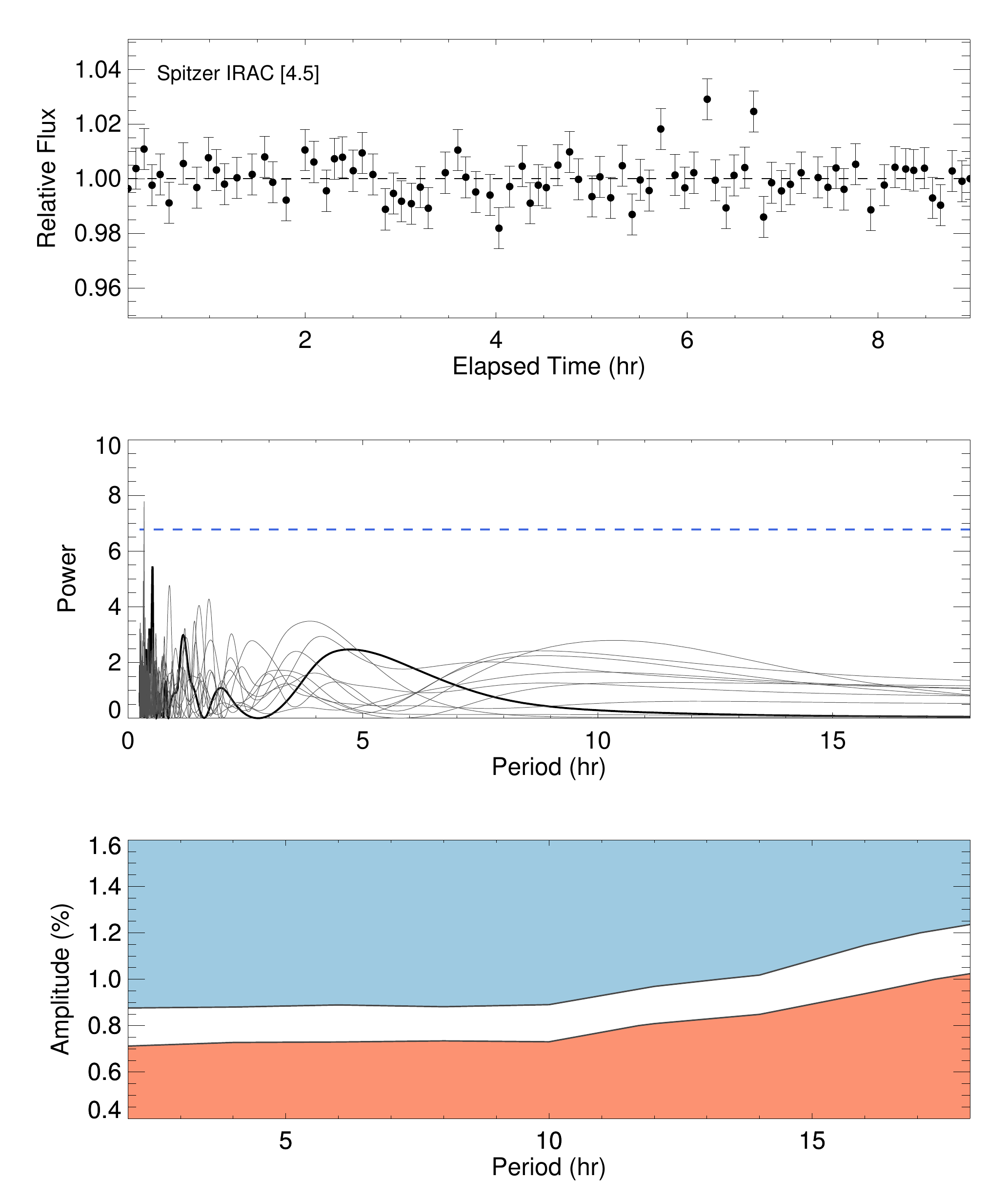}
	\caption{ The top panel shows the normalised, pixel phase corrected lightcurve of SDSS1110. The middle panel shows the periodogram of SDSS1110 (thick black line) as well as the periodograms of several other reference stars in the field. The $1\%$ FAP value is plotted in blue. The bottom panel shows the sensitivity of the observation as a function of amplitude and period. The blue area represents amplitudes detectable by the pipeline as a function of period (FAP $<1\%$), the white area shows amplitudes marginally detectable ($1\%<$ FAP $<5\%$), and the orange area shows amplitudes not detectable (FAP $>5\%$). For periods $<18~$hr, we place an upper limit of $1.25\%$ on the variability amplitude of SDSS1110.  }
	\label{fig:lcurveSDSS1110}
\end{figure}

\subsection{SDSS1110}
The light curve of SDSS1110 (top panel of Figure \ref{fig:lcurveSDSS1110}) does not display any obvious trends, and our periodogram analysis (middle panel) confirms this. 
To determine the sensitivity of our observation, we inject simulated sinusoidal curves into random permutations of our SDSS1110 lightcurve. The simulated sine curves have peak-to-peak amplitudes ranging from $0.4-1.6\%$ and periods of $2-18~$hr, with randomly assigned phase shifts. Each simulated lightcurve is put through our periodogram analysis, which allows us to produce a sensitivity plot, shown in the bottom panel of Figure  \ref{fig:lcurveSDSS1110}. The blue region corresponds to periods and amplitudes detected with a FAP $<1\%$, the white region corresponds to those detected with $1\%<\mathrm{FAP}<5\%$, and the orange region corresponds to those with FAP$>5\%$. For periods $<18~$hr, an upper limit of $1.25\%$ is placed on the variability amplitude of SDSS1110. Considering only periods $<10~$hr, as done by \citet{Metchev2015a}, we place an upper limit  of $0.9\%$ on the variability amplitude. However, since we expect that young brown dwarfs will rotate more slowly due to conservation of angular momentum, the limit based on periods $<18~$ hr is more robust. 

The photometric noise measured for SDSS1110 is comparable to the noise measured for comparison stars of similar brightness, and thus we do not find evidence for stochastic or aperiodic variability. \added{We additionally check the periodogram of the unbinned lightcurve to search for evidence of very short period ($<1~$hr) deuterium pulsations proposed by \citet{Palla2005}. The periodogram does not display significant peaks at these short periods. A photometric variability survey of late-M brown dwarfs with $T_{\mathrm{eff}}>2400~$K and ages of $1-10~$Myr  concluded that pulsations cannot grow to observable amplitudes in these objects \citep{Cody2014}. The absence of  short period pulsations detected in the light curve of SDSS1110 suggests that this conclusion may extend to even cooler ($T_{\mathrm{eff}}\sim900-1300~$K) brown dwarfs, however a larger sample will be needed to robustly explore this possibility. Deuterium pulsations are not expected to occur in objects with masses over the deuterium burning limit at the age of AB Doradus so would not be expected to occur in W0047 and 2M2244.} 

\section{The Inclination Angles of W0047 and 2M2244}\label{sec:inclination}

\begin{table}
\centering
\caption{Calculated {effective temperatures, $\log (g)$,} rotational velocities, radial velocities, periods, {$[3.6~\mu m]$ peak-to-peak variability amplitudes} and inclination angles for W0047 and 2M2244. }
\label{tab:results}
\begin{tabular*}{\columnwidth}{l@{\extracolsep{\fill}}ll}
\hline \hline
Parameter                		& W0047                			& 2M2244              \\[4pt]
\hline 
$v \sin i$ (kms$^{-1})$ 		& $9.8\pm0.3$  			& $14.3_{-1.5}^{+1.3}$             \\[2pt]
RV (kms$^{-1})$ 			& $-19.8 _{-0.2}^{+0.1}$  		& $-16.0_{-0.9}^{+0.8}~$           \\[2pt]
P (hr)                 			& $16.4 \pm 0.2$    			& $11.0 \pm 2.0$    \\[2pt]
$[3.6~\mu m]$ Amp ($\%$)              		& $1.07 \pm 0.04$    			& $0.8 \pm 0.2$    \\[2pt]
$R$ ($R_{Jup})$        		& $1.3\pm0.04$         		& $1.29 \pm 0.03$     \\[2pt]
$i$                    				& $85 ^{+5\circ}_{-9} $ 		& $76 ^{+14\circ}_{-20}$ \\[2pt]
\hline 
\end{tabular*}
\end{table}

With measured values for $v\sin i$ and the rotation period, $P$, in hand, an assumption of radius allows us to determine the angle of inclination, $i$.
We assume that the brown dwarf rotates as a rigid sphere. However, this is not strictly true. The rotational period of Jupiter, as measured by magnetic fields originating in the core is $9^h55^m40^3$, whereas the period measured using features rotating along the equator is $9^h50^m30^s$, a difference of $5$ minutes \citep{Allen}. Since rotational periods as measured from photometric variability in general have much larger uncertainties, the rigid body assumption is reasonable for our analysis. Thus, the equatorial rotation velocity, $v$, is given by $v=2\pi R / P$, where $R$ is the radius of the brown dwarf and $P$ is its rotation period. \added{We use the radii calculated from evolutionary models in Section \ref{sec:props}.}

Monte Carlo analysis was used to determine the inclination, $i$, for each target, using {the posterior $v\sin i$ distributions obtained from our MCMC analysis.} {For the period and radius we draw samples from a guassian distributed sample with a width given by the reported errors.} The radius estimates, rotational and radial velocities, periods, and resulting inclinations are shown in Table \ref{tab:results}.

\begin{figure}
	\centering
		\includegraphics[scale=0.5]{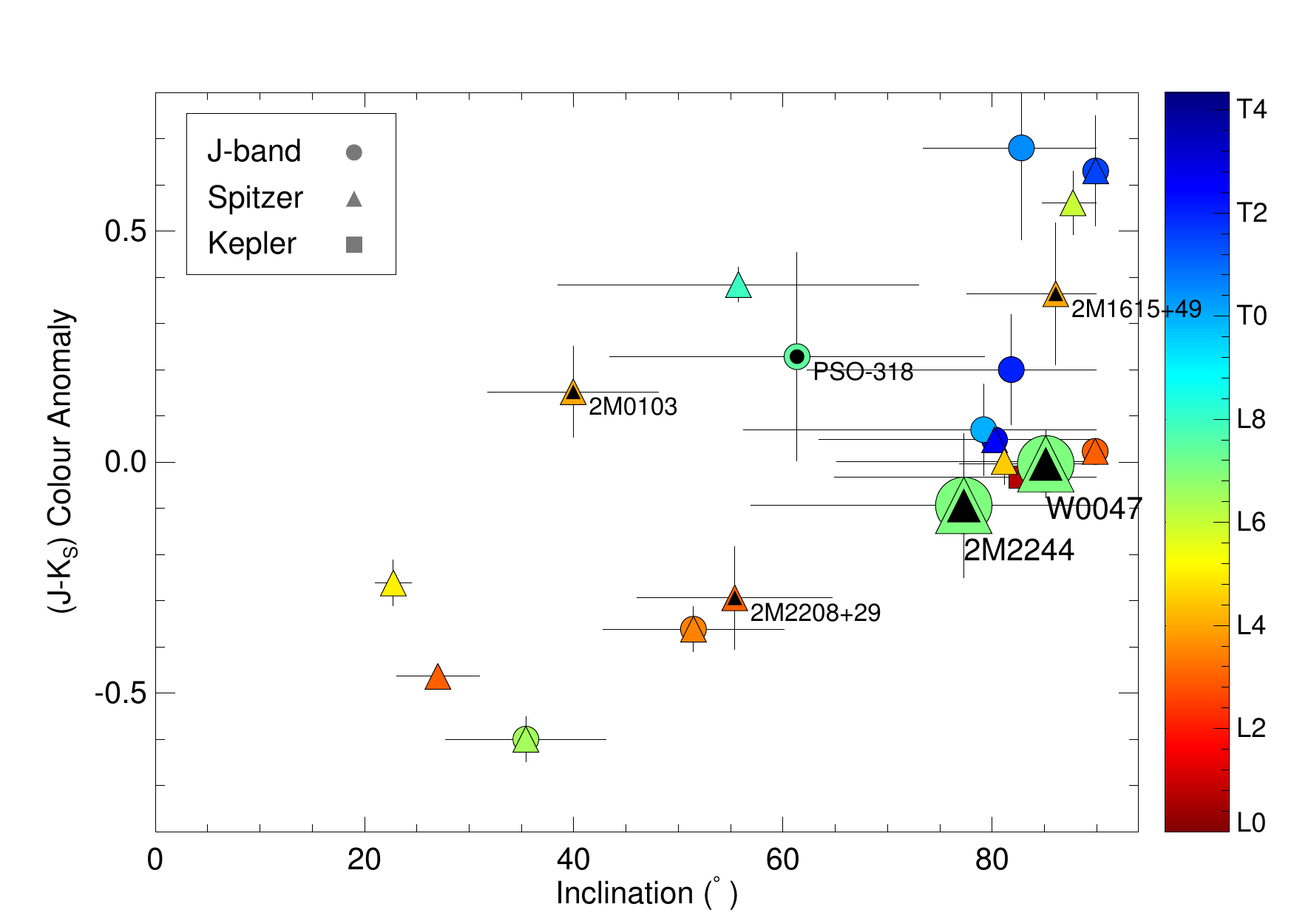}
	\caption{ $(J-K_S)_{\mathrm{2MASS}}$ colour anomaly plotted against the inclination angle for variable brown dwarfs. Black insets denote low-gravity brown dwarfs. In addition to 2M2244 and W0047, the low-gravity variable objects shown in this plot are 2M0103+19 (L4), 2M1615+49 (L4), PSO-318 (L7.5) and 2M2208+29 (L3). Inclination data for 2M2244 and W0047 are calculated in this paper, inclination data for other objects are from \citet{Vos2017}.}
	\label{fig:inc_col}
\end{figure}

We find an inclination angle of $85 ^{+5\circ}_{-9}$ for W0047, so this object is viewed {nearly} equator-on. This inclination is significantly larger than the $33^{+5\circ}_{-8}$ calculated by \citet{Lew2016}. This is as a result of both our longer period and larger $v \sin i$ measurements. The inclination angle of 2M2244 is found to be $76 ^{+14\circ}_{-20}$, {which is} similar to that of W0047. Considering their remarkably similar colours, spectra and inclination angles, the results are consistent with the idea that atmospheric appearance is influenced by viewing angle rather than rotation period or variability properties. Figure \ref{fig:inc_col} shows $(J-K_S)_{\mathrm{2MASS}}$ colour anomaly plotted against the inclination angle for variable brown dwarfs with our results for W0047 and 2M2244 overplotted \citep{Vos2017}. {The colour anomaly of each object is defined as the median $(J-K_S)_{\mathrm{2MASS}}$ colour for the spectral type and gravity flag of that object subtracted from its $(J-K_S)_{\mathrm{2MASS}}$ colour. Thus, positive and negative values of colour anomaly refer to objects that are redder and bluer than the corresponding median colour for their spectral types and gravity flags. } Median colours for L0 - T6 field objects {and their uncertainties} were taken from \citet{Schmidt2010}. \citet{Liu2016} provide linear relations between spectral type and absolute magnitude for {VL-G}  and {INT-G}  brown dwarfs, and these were used to calculate the median colours for the intermediate and low-gravity objects. {Since W0047 and 2M2244 have nearly identical spectra \citep{Gizis2015}, we treat them both as L7 INT-G objects, and apply the same colour anomaly correction to both.} {The error bars for these objects are simply the $J$ and $K_S$ magnitude uncertainties combined. Our estimate of the median colour of low-gravity objects is limited by the low number of such objects known. As more of these objects are discovered this median colour will become more accurate.} \citet{Vos2017} find that the correlation between near-infrared colour anomaly and inclination angle of field brown dwarfs is  statistically significant at the $99\%$ level. Variable brown dwarfs viewed equator-on appear redder {than} the median while objects closer to pole-on are bluer than the median. This figure is updated in Figure \ref{fig:inc_col}.  W0047, 2M2244 and the low-gravity objects 2M0103+19, 2M1615+49, PSO-318 and 2M2208+29 may follow this trend, {although more inclination data for young dwarfs are needed to fully explore this possibility. }
This may be explained if clouds are inhomogeneously distributed in latitude or if grain size and cloud thickness vary in latitude. If thicker or large-grained clouds are situated predominantly at the equator, while thinner or small-grained clouds are situated at the poles then we would expect to observe objects with $i\sim90^{\circ}$ to be redder than the median and objects with lower inclination angles to be bluer than the median. { The addition of more inclination data for brown dwarfs is likely to reveal the physical origin of the correlation seen in Figure \ref{fig:inc_col}.}

\begin{figure}
	\centering
		\includegraphics[scale=0.5]{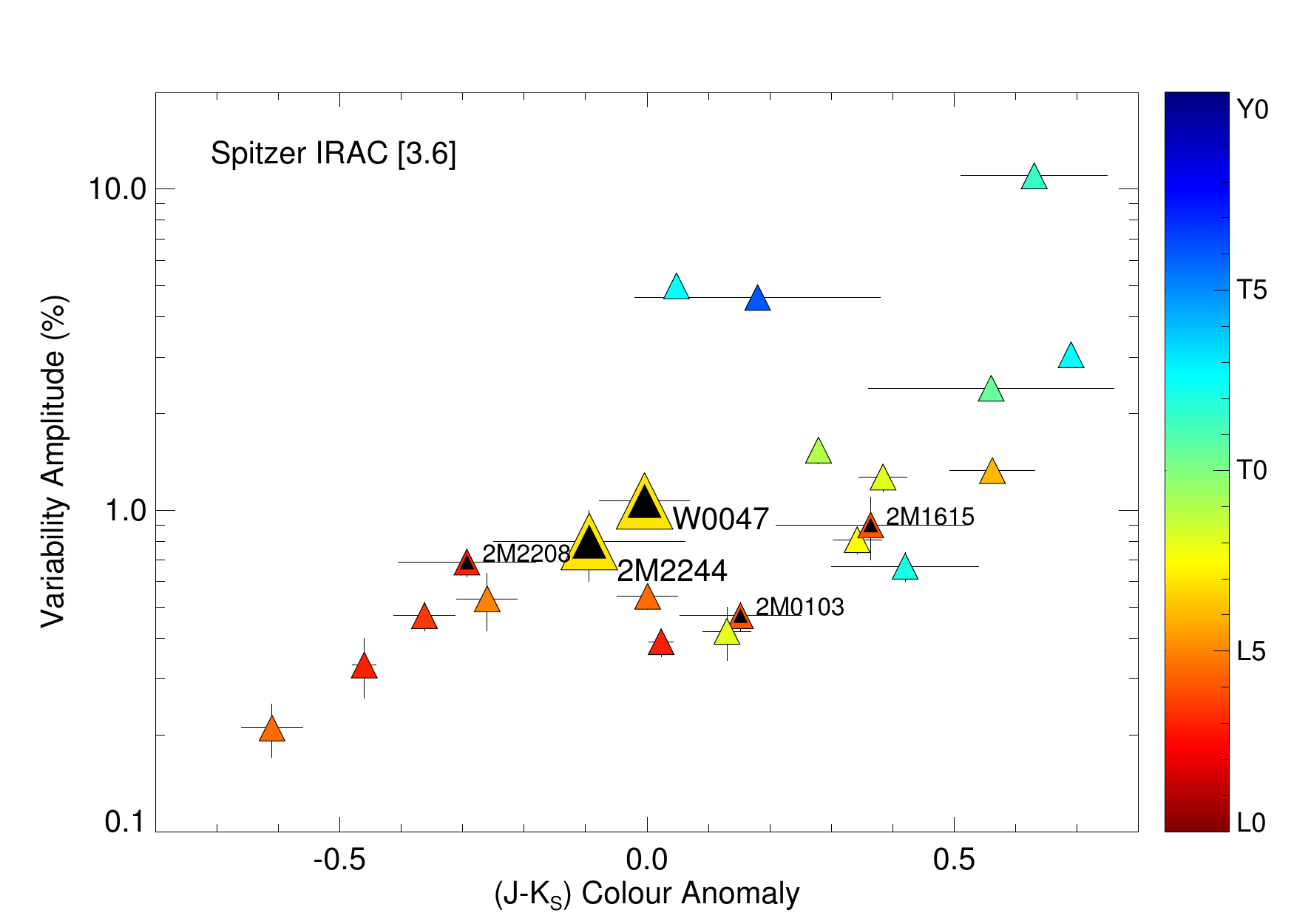}
	\caption{ \textit{Spitzer} $[3.6~\mu$m$]$ variability amplitude plotted against $(J-K_S)_{\mathrm{2MASS}}$ colour anomaly for variable brown dwarfs. Data for 2M2244 and W0047 are presented in this paper, data for other objects are from \citet{Vos2017}.}
	\label{fig:amp_col}
\end{figure}

{\citet{Vos2017} also find a relation between the colour anomaly of an object and its variability amplitude, where objects that are redder than the median for their spectral type and gravity class tend to have higher variability amplitudes. Figure \ref{fig:amp_col} shows an updated version of this plot, showing that W0047 and 2M2244 are also consistent with this trend.}

\section{Conclusions}

We have obtained \textit{Spitzer} [$3.6~\mu m$] photometric monitoring for two young free-floating objects, W0047, 2M2244 and \textit{Spitzer} [$4.5~\mu m$] monitoring of SDSS1110 {as well as $J$-band monitoring of 2M2244}. Additionally, we obtain NIRSPEC N-7 spectra of W0047 and 2M2244. We detect variability in the two late-L, low mass dwarfs W0047 and 2M2244.
MCMC analysis of the \textit{Spitzer} [$3.6~\mu m$] lightcurve of 2M2244 gives a period of $11\pm2~$hr and a peak to trough amplitude of $0.8\pm0.2\%$. {We detect significant ($\sim3\%$) $J$-band variability in 2M2244.}
We find a period of $16.4 \pm 0.2~$hr for W0047 and an amplitude of $1.07\pm0.04\%$. Variability is not observed in the T5.5 object SDSS1110 during an $8.5~$hr observation. For periods $<18~$hr, we place an upper limit of $1.25\%$ on the variability amplitude of SDSS1110.

With a peak to trough amplitude of $1.07\pm0.04\%$ for W0047, this is among the highest \textit{Spitzer} [$3.6~\mu$m] variability amplitudes detected. This variability detection adds to a growing number of young, L-type objects that display high amplitude variability, suggesting that this correlation may extend into the late-L spectral types \citep{Metchev2015a, Biller2015, Lew2016}.

 The $v \sin i$ of both targets is determined using NIRSPEC-7 high dispersion spectra, finding $v \sin i = 14.3_{-1.5}^{+1.3}~$km s$^{-1}$for 2M2244 and $v \sin i = 9.8\pm0.3~$km s$^{-1}$for W0047.
Assuming rigid sphere rotation and using expected radii from evolutionary models, we find that both objects are close to equator-on, with inclination angles of $85 ^{+5\circ}_{-9}$ and $76 ^{+14\circ}_{-21}$ for W0047 and 2M2244 respectively.
Their remarkably similar colours, spectral appearance and inclination angles is consistent with the possibility that viewing angle shapes the observed spectrum of a brown dwarf or giant exoplanet.

\section*{Acknowledgements}

JV acknowledges the support of the University of Edinburgh via the Principal's Career Development Scholarship. KA acknowledges support from the Isaac J. Tressler Fund for Astronomy at Bucknell University. BB gratefully acknowledges support from STFC grant ST/M001229/1. \added{We thank Michael Kotson for assistance with the evolutionary models.}
This work is based [in part] on observations made with the \textit{Spitzer Space Telescope}, which is operated by the Jet Propulsion Laboratory, California Institute of Technology under a contract with NASA. Support for this work was provided by NASA through an award issued by JPL/Caltech.
This work was supported by a NASA Keck PI Data Award, administered by the NASA Exoplanet Science Institute. Data presented herein were obtained at the W. M. Keck Observatory from telescope time allocated to the National Aeronautics and Space Administration through the agency's scientific partnership with the California Institute of Technology and the University of California. The Observatory was made possible by the generous financial support of the W. M. Keck Foundation. UKIRT is owned by the University of Hawaii (UH) and operated by the UH Institute for Astronomy; operations are enabled through the cooperation of the East Asian Observatory. When the data reported here were acquired, UKIRT was supported by NASA and operated under an agreement among the University of Hawaii, the University of Arizona, and Lockheed Martin Advanced Technology Center.
The authors wish to recognise and acknowledge the very significant cultural role and reverence that the summit of Mauna Kea has always had within the indigenous Hawaiian community. We are most fortunate to have the opportunity to conduct observations from this mountain.



\bibliographystyle{mnras}
\bibliography{Full} 









\bsp	
\label{lastpage}
\end{document}